\PassOptionsToPackage{table}{xcolor}
\documentclass[runningheads,final]{llncs}

\newif\ifanonymous
\anonymousfalse

\usepackage{array}
\usepackage{amssymb}
\usepackage{booktabs}
\usepackage{siunitx}
\usepackage{xcolor}
\definecolor{lightergray}{rgb}{0.97, 0.97, 0.97}
\definecolor{darkred}{rgb}{.6, 0, 0}
\definecolor{darkgreen}{rgb}{0,.4, 0}
\definecolor{darkblue}{rgb}{0, 0,.6}
\definecolor{darkergreen}{rgb}{0,.8,0}
\usepackage{tabularx}
\usepackage{subcaption}
\usepackage[
  colorlinks=true,
  linkcolor=black,
  citecolor=darkgreen,
  filecolor=darkblue,
  urlcolor=darkblue,
  bookmarksopen=true,
  bookmarksopenlevel=3,
  plainpages=false,
  breaklinks=true
]{hyperref}
\usepackage[
  colorinlistoftodos,
  obeyFinal,
]{todonotes}
\usepackage[
  capitalise,
  nameinlink
]{cleveref}
 
\usepackage{xurl}

\usepackage{ellipsis, ragged2e, marginnote}
\usepackage[tracking=true]{microtype}
\DeclareMicrotypeSet*[tracking]{my}{ font = */*/*/sc/* }
\SetTracking{encoding = *, shape = sc}{45}

\begin{document}
\title{Probing for Passwords -- \\Privacy Implications of SSIDs in Probe Requests}

\ifanonymous
\else
    \author{%
      Johanna Ansohn McDougall
      \and Christian Burkert
      \and Daniel Demmler
      \and Monina Schwarz
      \and Vincent Hubbe
      \and Hannes Federrath
    }
    \authorrunning{J. Ansohn McDougall et al.}
    %
    \institute{%
    University of Hamburg, Germany\\
    \email{\{johanna.ansohn.mcdougall, christian.burkert, daniel.demmler, monina.schwarz, vincent.hubbe,  hannes.federrath\}@uni-hamburg.de}\\
    }
\fi

\maketitle

\begin{tikzpicture}[remember picture, overlay]
  \node[font=\sffamily\normalsize, yshift=-1cm, text centered, text width=\paperwidth, anchor=north west] at (current page.north west) {%
The final publication is available at Springer via \url{https://doi.org/10.1007/978-3-031-09234-3_19}
  };
\end{tikzpicture}

\begin{abstract}
  Probe requests help mobile devices discover active Wi-Fi networks. They often contain a multitude of data that can be used to identify and track devices and thereby their users.
  The past years have been a cat-and-mouse game of improving fingerprinting and introducing countermeasures against fingerprinting.
  
  This paper analyses the content of probe requests sent by mobile devices and operating systems in a field experiment. In it, we discover that users (probably by accident) input a wealth of data into the SSID field and find passwords, e-mail addresses, names and holiday locations. With these findings we underline that probe requests should be considered sensitive data and be well protected.
  To preserve user privacy, we suggest and evaluate a privacy-friendly hash-based construction of probe requests and improved user controls.
\end{abstract}


\section{Introduction}

To establish a Wi-Fi connection, mobile devices can transmit so-called \emph{probe requests} to receive information about nearby Wi-Fi networks. An access point observing a probe request is led to reply with a probe response, thereby initiating a connection between both devices.
While probe requests are used to establish a connection between a mobile device and an AP, they also serve as a means to track, trilaterate and identify devices for attackers who passively sniff network traffic. They can contain identifying information about the device owner depending on the age of the device and its OS. 
One of those is the preferred network list~(PNL), which contains networks identified by their so called Service Set Identifier~(SSIDs).
Around 23\,\% of the probe requests contain SSIDs of networks the devices were connected to in the past, according to our measurements.
There exist online mapping services like WiGLE\footnote{\url{https://www.wigle.net}}, which provide information about geographical locations where SSIDs have been observed.
A casual observation of the networks available in any given residential area returns a multitude of personalised, often descriptive SSIDs used for private networks.
Therefore, a query for an SSID might reveal home or work addresses, or other visited locations where users connected to Wi-Fi, and can thereby reveal very personal information about them.

Another application in which probe requests are frequently used is tracking of devices in stores or cities: as probe requests are sent rather frequently, they can be used to trilaterate the location of a device with an accuracy of up to 1.5\,metres~\cite{redondi2018building}.
Trilateration can also be used to follow the movements of a device and thereby its user over a longer period of time, and track them through a store or city~\cite{eckernfoerdeTrackt}. This is in fact employed in 23\,\% of the stores already~\cite{Acar.2018}.
Companies and cities that conduct Wi-Fi tracking take the legal position that only the MAC address contained in probe request is considered personal data according to GDPR Article~4(1)~\cite{bundestagTracking,dsgvo}, which protects personal data from unlawful collection and processing.
They therefore maintain that if the MAC address is anonymised before storage, the collection and evaluation of probe requests is GDPR compliant~\cite{expocloud}.
The randomisation of MAC addresses mitigates linkability via this element.
Instead, we focus on looking at what privacy risks originate from probe requests related to the list of SSIDs stored in the PNL\@.
We provide empirical evidence that probe requests should also be considered personal data on the basis of their SSID field, which we find can even contain directly identifying information.
We hope to thereby stress the need for a more thorough legal evaluation. We additionally propose changes to the handling of SSID field and mobile OS behaviour to enhance the privacy of users and decrease their trackability to passive sniffers. 

\bigskip\noindent
To this end, we contribute the following:
\begin{itemize}
    \item We conduct a field experiment in a German city, recording probe requests of passersby.
    \item We evaluate their content, with special regard for SSIDs and identifying information.
    \item We summarise the state of probe requests for different OS versions.
    \item We propose a hashing of non-wildcard SSIDs in probe request to protect their confidentiality against passive observers.
    \item We propose changes to the UI design of Wi-Fi selection and PNL management.
\end{itemize}

This paper is structured as follows: in the next section, we first  provide a background on network discovery and privacy implications of MAC addresses. Here, we also compare the privacy features of various Android and iOS versions. Thereafter, we present related work in \Cref{sec:related work}. \Cref{sec:experimental_setup} explains the experimental setup and our handling of ethical and privacy concerns.
We then present the results of our data analysis in \Cref{sec:data analysis}.
\Cref{sec:Privacy Enhancement for the use of Hidden Networks} proposes mitigation approaches on both protocol and user interface level. 
In \Cref{discussion and limitations}, we discuss the findings. Finally, \Cref{conclusion} concludes the paper.

\section{Background}
\label{sec:background}

In this section we define the underlying technological background of our work.

\subsection{Network Discovery in 802.11}

To establish a Wi-Fi connection between a mobile device and an access point~(AP), both devices have to discover each other; either via active or passive discovery:

In a passive discovery, an AP advertises itself by sending out \emph{beacons} containing its SSID, MAC address, the cipher suites it supports and a few other elements~\cite{OWE-rfc}. These beacons are sent at an interval of approximately every 100\,ms~\cite{TalkativePhones}, and mobile devices can respond with Wi-Fi association frames. 

In an active discovery, mobile devices broadcast \emph{probe requests} to find APs they have previously associated with. 
Active discovery is also required to connect to so-called hidden networks, for which the AP does not advertise the network, i.e., does not send out beacons.
Probe requests sent by most modern devices are typically broadcast and contain the empty wildcard in the SSID field.
APs receiving a probe request respond with a \emph{probe response} directed at the sender of the probe request. The probe response contains the~SSID of the~AP and additional information like supported rates and various capabilities. 

The reason both active and passive discovery mechanisms are used is that while APs advertise themselves constantly, scanning for beacons can be rather energy consuming and slow.
Additionally, a mobile device scanning for beacons on one channel with a certain frequency might miss beacons sent on another channel.
A device actively probing for APs just has to turn on the Wi-Fi radio until it receives the probe response, which typically takes only a few milliseconds~\cite{TalkativePhones}.
On the other hand, active discovery requires the transmission of packets containing information about the mobile device.
While probe requests sent by devices running older OS might contain SSIDs of one or more APs the device has previously been connected to, newer devices transmit only the SSIDs of hidden networks to improve user privacy and make the device less traceable (cf. \cref{sec:Differences between Android and iOS versions}).
Additionally, they omit the real MAC address of the device, instead sending a randomised MAC address.

Probe Requests are sent in bursts, every burst containing several probe requests sent via some or all of the 14 channels of the 2.4\,GHz spectrum (and additionally the 5\,GHz spectrum if applicable) within a short time span of just a few milliseconds. Whether MAC address randomisation is employed or not, all packets in a burst are sent from the same MAC address.

\begin{figure}
    \centering
    \includegraphics[scale=0.034]{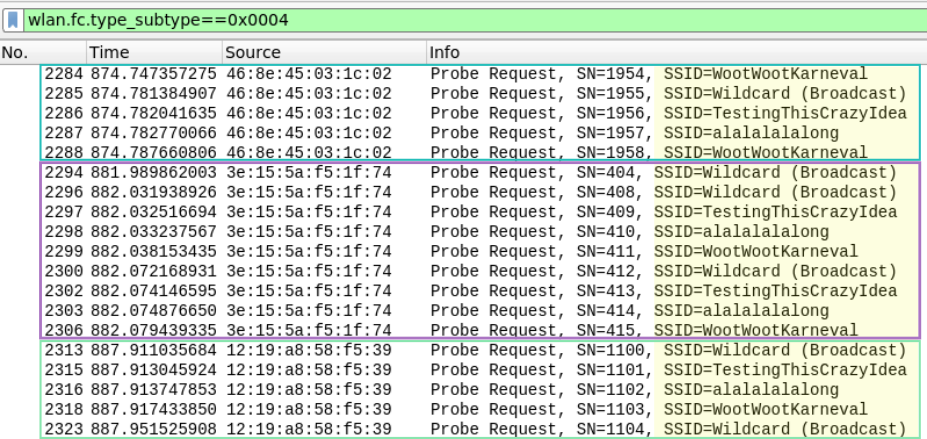}
    \caption{Three bursts of probe requests sent from the same device. Three different SSIDs and the wildcard SSID, an empty string, are broadcast. Note that the starting sequence number~(SN) in the info field is randomised per burst as well.}
    \label{fig:bursts}
\end{figure} 

\subsection{Privacy Implications}

A MAC address consists of \SI{6}{bytes} typically represented in hexadecimal notation, separated by colons, e.\,g. \texttt{01:23:45:ab:cd:ef}.
The first three bytes are called the Organizationally Unique Identifier~(OUI) and are typically assigned to the manufacturer of the devices.
The last three bytes identify the Network Interface Controller~(NIC), produced and assigned by the manufacturer.
The~OUI contains additional information encoded in the two least significant bits (U/L and I/G) of the most significant byte (\texttt{01} in our example):
The I/G-bit is the least significant bit and specifies whether the recipient is unicast or a multicast.
The second-least significant bit, the U/L bit, clarifies whether the address is locally or globally administered, with a globally administered address being a unique identifier for the physical device, while a locally administered address temporarily overwrites the unique global one in software~\cite{MACaddrRandRFC}.
Older devices use their universal address to broadcast probe request, which makes them easily trackable.
To protect the privacy of users and prevent device tracking, probe requests are often sent from locally administered addresses,  employing a technique called MAC address randomisation.
Here, the MAC address commonly changes between two bursts, such that each burst will be sent from a new, random MAC address.
This behaviour was first introduced in iOS~8 in~2014 \cite{MARAndWhenItSucceeds} and in Android 8~\cite{android_macrandom} in~2017, albeit the first implementations suffered from information leaks: it was often possible to track devices despite the use of MAC address randomisation~\cite{Vanhoef-RandomisationNotEnough}, for example by the SSIDs they contained.  
If SSIDs are present in a probe request, either all of them or a subset is contained in a burst, with every packet requesting one SSID.
\cref{fig:bursts} shows the capture of three bursts of probe requests sent from the same device employing MAC address randomisation but transmitting SSIDs.

While omitting SSIDs and employing MAC address randomisation renders a device less trackable, other fields included in probe requests can spoil the effect: if the sequence number~(SN) is not randomised, it is trivial to still follow a device over time. Therefore, a lot of devices randomise their sequence number with the start of every burst, as can also be observed in \cref{fig:bursts}. 

The newer a device and its OS is, the more information is omitted and fields randomised in the probe requests.
All the same, various papers still describe how even modern devices can be fingerprinted due to other information contained in them, e.\,g. in the Information Elements~(IE): These non-mandatory parameters contain information on supported rates, network capabilities, and more. Combining the IE parameters, the signal strength and, in some cases, the sequence number, allows to fingerprint individual devices despite MAC address randomisation.~\cite{Vanhoef-RandomisationNotEnough,tan2021efficient}  
While efforts are made to reduce the fingerprint of modern devices, the owners of older devices that don't receive patches introducing MAC address randomisation, sequence number randomisation and SSID omission can easily be tracked.

\subsection{Differences between Android and iOS versions}\label{sec:Differences between Android and iOS versions}

\Cref{tab:sec_features} shows the differences between iOS and Android in supporting Wi-Fi-related privacy features.
We compare iOS versions 8, 10, 14 and 15 and Android~8 to 12.
Their combined market share comprises about 90\,\% of the devices \cite{android_version_share,ios_version_share}, which makes them representative and provides a good overview over the changes within the last years. In the following, we elaborate on the various features comprised in \cref{tab:sec_features}.


\begin{table}[bph]
\newcommand\y{\checkmark}
\newcommand\no{-}
\newcolumntype{C}{>{\centering\arraybackslash}p{2.3em}}
\caption{Privacy features for probe requests in different mobile OSs.}%
\label{tab:sec_features}
\begin{tabularx}{\textwidth}{%
  X|*{4}{C}|*{5}{C}
}%
  \toprule
  & \multicolumn{4}{c|}{Apple iOS} &  \multicolumn{5}{c}{Android}  \tabularnewline
  & 8 &  10 & 14 & 15 &  8 & 9 & 10 & 11 & 12 \tabularnewline
  \midrule
  Market Share in \% & \mbox{$<0.1$} & 1.0 & 35.9 & 53.4 & 10.2 & 13.5 & 27.0 & 35.4 & 1.9 \\
  \midrule
  Randomised MAC \dots &  &  &  &  &  &  &  &  &  \\
   - while probing  &  \y &  \y &  \y &  \y &  \y &  \y &  \y &  \y &  \y \tabularnewline

  - per connected  SSID &  \no &  \no &  \y &  \y & \no  & (\no)*  &  \y &  \y &  \y \\

  - after resetting settings &  \no &  \no &  \y &  \y &  \no &   \no &  (\no)* &  (\no)* &  (\no)*   \\

  \midrule
  New random MAC after ...&  \no &  \no &  \no &  {\small{6w}} &  \no &  \no &  \no &  (\no)$^\dagger$ &  (\no)$^\dagger$ \\

  Private Address by default &  \no &  \no &  \y &  \y &  \no &  \no &  \y &  \y &  \y \\

  Modify distant Network &   \no &  \no &  \no &  \no &  \y &  \y &  \y &  \y &  \y \\
  
  Manually added == hidden & \multicolumn{4}{c|}{\tiny{Automatic detection of hidden}} &  \y  &  \no &  \no &  \no &  \no\\
  
  Probe with SSID & \multicolumn{4}{c|}{\tiny{Only if hidden detected}} & \multicolumn{1}{m{2.2em}}{\tiny{if man. added}}   &  \multicolumn{4}{c}{\tiny{If explicitly declared hidden}}  \\

  \bottomrule
  \multicolumn{10}{p{\linewidth}}{\scriptsize
    *: Only choosable via Developer Options\newline
    $\dagger$: If use of non-persistent MAC is chosen via Developer Options, a new MAC is set (a) for every new connection establishment (b) every 24 hours, unless a connection is still established or (c) if both the DHCP lease has expired and the device has been disconnected for 4 hours\newline
  }
\end{tabularx}
\end{table}

All listed versions use MAC address randomisation while probing \cite{martin2017study,android_macrandom,MACaddrRandRFC}. 
In Android 9 devices, users can choose via Developer Options whether a randomised MAC address should be used while connected. Since Android 10 and iOS 14, private addresses are used by default: They all employ persistent random MAC addresses while connected.
Starting with Android~11, one can choose via Developer Options to use non-persistent randomisation per stored SSID during connection. If non-persistent MAC addresses are used in Android, the MAC address is re-randomised either (a) with every new connection establishment, (b) every 24 hours, but without disrupting the network connection to switch to a new MAC address or (c) if the DHCP lease has expired and the device has been disconnected for at least 4 hours. 
Nevertheless, up to this date, there are no Android devices that automatically receive a new random MAC address after a certain amount of time by default, without having to modify the Developer Options. Additionally, with persistent private MAC addresses in Android, the default behaviour is to persist a MAC address per SSID even after resetting network settings. This is different in iOS: Starting with iOS 14, removing and adding a network again causes a reset of the network persistent address. In iOS 15, the devices additionally receive a new address when not connected for more than 6 weeks.\cite{applePrivateWifi,AndroidMacBehavior,MACaddrRandRFC}

All of the listed Android versions offer to remove any saved network at any given time. This is not the case in iOS: Here, a network can only be removed from the device while in physical proximity of it or by modifying the iCloud Keychain from a MacBook. Without access to a MacBook or physical proximity to the network, it can not be removed without resetting the entire network settings.\cite{removeProbesFromiPhone}

When adding a network manually, iOS verifies whether it is a hidden network or not, whereas Android 8 (and earlier) automatically assumes that manually added networks are hidden networks. Therefore, if a network was manually added, Android 8 devices send the SSID in probe requests, while newer Android versions only do so if the added networks were explicitly declared hidden (cf. \cref{fig:hidden_message} in \cref{sec:evaluation}). In iOS, the SSID is only used in probe requests if the network is detected to be a hidden network.\cite{PlatformSecurity2021,AndroidScanBehaviour}

\section{Related Work}\label{sec:related work}

In 2014, Cunche et al. showed how to link various devices by their transmitted SSIDs and inferred relationships between users~\cite{cuncheLinkingWirelessDevices2014}.
This work was published before MAC address randomisation was deployed and free transmittal of SSIDs the typical means of network discovery. The authors propose the use of a geolocation-based service discovery instead of active discovery via probe requests. Later in 2014, MAC address randomisation was first discussed \cite{MACaddrRandRFC} and subsequently tested and published \cite{bernardosWiFiInternetConnectivity2015}. It was meant as a means to increase privacy, but since it lacked standardisation, all implementations were vulnerable to attacks \cite{Vanhoef-RandomisationNotEnough}. Since then, extensive work has been published on probe requests, MAC address randomisation and fingerprinting devices despite MAC address randomisation:
2015, Freudiger et al. gave an overview over the amount of probe requests sent by different devices and analysed the effectiveness of the MAC address randomisation employed in different devices~\cite{TalkativePhones}. On a positive note, Freudiger pointed out that recent mobile operating systems only probe for SSIDs of hidden networks.
In another influential publication, Vanhoef et al. investigated how well devices can be tracked by combining various fields in probe requests~\cite{Vanhoef-RandomisationNotEnough}. They also present two attacks that can be used to reveal the real MAC address of a device and summarise that MAC address randomisation is insufficient to impede tracking.

Various other papers in the field attempt to associate probe requests from randomised MAC addresses: Gu et al. use deep learning methods and suggest to encrypt probe requests using the symmetric stream cipher ChaCha20 to protect them from attackers~\cite{ProbeReqIdentification}.
Tan et al. use minimum-cost flow optimisation to associate frames and reach an accuracy of more than 80\,\%~\cite{tan2021efficient}. As the use of hidden networks and the amount of devices broadcasting SSIDs are decreasing, both papers put only a minor focus on the transmitted SSIDs.

In 2019, Dageli{\'c} et al. \cite{dagelic2019location} observe the occurrence of SSIDs in probe requests at a music festival between 2014 and 2018 and present how easy devices are trackable via probe requests if the devices are fingerprintable.
Over the years, the number of MAC addresses they observe increases while the number of SSIDs decreases. They conclude that the use of MAC address randomisation is increasing, as is the number of probe requests that contain the empty wildcard SSID. 

An attempt at localisation of criminal groups via probe requests was published by Zhao et al. in 2019~\cite{zhao2019localization}.
They build a database of SSIDs like WiGLE and monitor probe requests in different locations in search for specific SSIDs. This methodology allows them to find and track devices belonging to a targeted group.

With respect to protecting the content of probe requests, Pang et al.~\cite{Pang07tryst:the} published an architecture called Tryst in 2007 to conceal confidential information during service discovery.
Tryst makes use of access control primitives using symmetric encryption, with which it reveals information to the correct access point while concealing all information not directed at it.
It remains unclear how exactly the various SSIDs present in the SEND primitive are concealed from everyone except for the intended recipient.
They underline the privacy risks of both APs transmitting SSIDs and mobile devices transmitting probe requests by analysing geoinformation on SSIDs collected in a 2004 data set: they find that about a quarter of the devices probe for SSIDs that uniquely appear in just one city.
While Pang et al. also perform geolocalisation of SSIDs like we do, to the best of our knowledge, there has been no publication analysing the content of SSIDs of probe requests as we do in this paper and neither one proposing hash-based anonymisation of probe requests to this date.

In the following, we first introduce the experiment and then strive to demonstrate the privacy implications of the use of such verbose devices on their users.

\section{Experimental Setup}
\label{sec:experimental_setup}

In this field experiment performed in November 2021, we recorded probe requests in a busy pedestrian zone in the centre of a German city, over the period of one hour, three times in total. 
We used six off-the-shelf antennae: three for channels 1, 6 and 11 in the \SI{2.4}{GHz}~spectrum and three for channels~36, 40 and 48 in the \SI{5}{GHz}~spectrum.
Since our particular focus lies on privacy violations arising from the information contained in the SSID field, we evaluate it with respect to the following: 
 
\begin{itemize}
    \item The amount of probe requests containing non-empty SSIDs.
    \item The amount of SSIDs sent per burst.
    \item Privacy implications of transmitted SSIDs: what potentially personal data can be gleaned from the data set?
    \item The use of MAC address randomisation.
\end{itemize}

We calculate an intersection between the data sets of two different days and remove all probe requests by devices that appear in both of them. That way, we strive to isolate the permanent devices in the vicinity of the measurement to have a clearer view on devices more likely to represent human passersby.

\subsection{Potential Ethical and Privacy Concerns}\label{sec:ethical and privacy concerns}

A modern smartphone might use MAC address randomisation and refrain from transmitting SSIDs and thereby protect the identity of its user and render itself less trackable.
Older devices are often less privacy sensitive, transmit their real MAC address and maybe even known SSIDs.
This data can be considered personal data and should therefore only be collected and stored with particular care for the device owner's privacy.
To ensure ethical data aggregation, we submitted our study for approval to the ethics committee of the Informatics department of the University of Hamburg under case number 002/2021. 
The steps taken to protect the peoples privacy as observed and in accordance with the ethics committee can be found in \cref{sec:Ethical collection of probe requests}.

\section{Data Analysis}
\label{sec:data analysis}

Our field data set contains \num{252242}~probe requests.
We found that overall, 23.2\,\% of the probe requests contained 
SSIDs.
Prior measurements done by Dageli{\'c}~\cite{dagelic2019location} between 2014 (46.7\,\%) and 2018 (12.9\,\%), and also Vanhoef~\cite{Vanhoef-RandomisationNotEnough} in 2016 (29.9\,\% to 36.4\,\%), revealed higher numbers in 2014 and 2016, from which  a decline is absolutely expected.
At the same time, the records of 2018 and our measurements do not match up.
One explanation might be, that while a measurement at a music festival might record the probe requests of younger people with more recent devices that already omit SSIDs in probe requests, our measurements were taken in the city centre of a touristic city around noon, where perhaps a larger percentage of people kept their (older) devices over a longer period of time.
Our numbers do however correlate with the market share of Android devices~\cite{android_version_share}:
10.2\,\% of Android devices use Android 8, and devices older than Android 8 amount up to 12\,\%.
In these devices, manually added networks are considered hidden networks~\cite{AndroidScanBehaviour} and they are therefore probed for with SSID.
Seeing that Android devices make up approximately 70\,\% of the market share, while iOS~devices make up around 29\,\%~\cite{comparison_ios_android}, the percentage of SSIDs in the data set is slightly higher than expected.

During our measurement, \num{116961}~probes~(46.4\,\%) were captured in the 2.4\,GHz spectrum, of which \num{28836}~(24.7\,\%) contained at least one SSID.
In the 5\,GHz spectrum, we recorded \num{135281}~probes~(53.6\,\%), of which \num{29653}~(21.9\,\%) contained an SSID.

\begin{table}[t]
    \centering%
    \caption{Distribution of the number of SSIDs per cluster.}%
    \label{tab:ssids_per_cluster}
 \begin{tabular}{l@{\hskip 2ex}*{9}{@{\hskip 1ex}r@{\hskip 1ex}}}%
  \toprule
  \# SSIDs & 1 & 2 & 3 & 4 & 5 & 6 & 7 & 8 & $>8$\\
  \midrule
   Share & 67.8\,\% & 8.2\,\% & 4.0\,\% & 6.5\,\% & 2.7\,\% & 2.2\,\% & 0.9\,\% & 6.6\,\% & 1.1\,\%\\
  \bottomrule
\end{tabular}
  
\end{table}

To prepare the probe requests for analysis, we first grouped all requests that were sent from the same MAC address within a period of four seconds into bursts.
We then grouped all bursts into clusters of bursts, likely belonging to a single device, if their PNL was equal.
We explicitly did not group requests into the same cluster if their PNL matched only partly to avoid misclassification of distinct devices with partly overlapping PNLs.
At the same time, a cluster with only one SSID might contain requests from distinct devices. As can be seen in \cref{tab:ssids_per_cluster}, 67.8\,\% of the bursts contain just one SSID, while the remaining 32.2\,\% contain more than one SSID and are unique enough to track devices with it.
This is considerably less than Vanhoef et al. recorded in 2016 \cite{Vanhoef-RandomisationNotEnough}: In their data set, 53\,\% to 64.8\,\% of the bursts contained a unique PNL. 

Of the probe requests containing an SSID, we identified at least 362~devices sending requests from multiple randomised MAC addresses. 542~devices used only one MAC address, and did not employ MAC address randomisation. 

In an additional evaluation, we found that the average amount of probe requests sent per unique MAC address was 4.8. This is, again, a legitimate decline in comparison to the capture by Dageli{\'c}~\cite{dagelic2019location} et al. in 2014 (24.1), 2015 (29.2) and 2017 (6.1) respectively, but is, again, higher than their 2018 count (2.6).
For packages including SSIDs, the average amount of probe requests sent per MAC address was 11.2, which confirms that these are likely older devices of which less employ MAC address randomisation.

In the following, we analyse the values contained in the SSID field of probe requests in order to estimate the privacy violations that occur in their commercial collection and analysis.

\subsection{SSID Contents}

As mentioned in \cref{sec:Differences between Android and iOS versions}, devices running Android 8 and lower treat manually added networks like hidden networks \cite{AndroidScanBehaviour}. We conjecture that a lot of the SSIDs in our record originate from users trying to set up a network connection manually by entering both SSID and password through the advanced network settings, and, apparently mistakenly, enter the wrong strings as the SSIDs.
The devices then retransmit the PNL with every probe burst.
This results in significant additional information for fingerprinting devices compared to the empty wildcard SSID that would be transmitted otherwise.

In the following, we elaborate on our findings of a manual, as well as automated inspection of the encountered SSIDs.

\subsubsection{Password Leaks in SSID Broadcasts.}\label{sec:pwleaks}

A small but significant amount of probe requests containing SSIDs potentially broadcast passwords in the SSID field: 
We identified that 11.8\,\% of the transmitted probe requests contain numeric strings with 16~digits or more, which are likely the initial passwords of popular German home routers (e.\,g., FritzBox or Telekom home router). 
This hypothesis is supported by various cases, in which the numeric strings follow a \texttt{PW:}, \texttt{WPA:} or \texttt{(WPA/WPA2:)}.
Also, we repeatedly found both a 16- or 20-digit string and in the same burst additionally the same string, but separated by a space, a dot or a comma every four digits (e.\,g.\ \texttt{1234567812345678} and \texttt{1234 5678 1234 5678}), which is a typical way of improving readability of the initial password on the router case. The multitude of similar spellings support our assumption that the users tried the same way of logging in multiple times with different spellings of the same credentials.

Leaking passwords in SSIDs is especially critical if, along with the password, the device also broadcasts the true SSID either correctly or with a mistype that can be used to infer the true SSID. Only 2.8\,\% of the transmitted SSIDs classified as probable passwords were the only entry in the corresponding PNL.
All other probable passwords were transmitted in bursts with other SSIDs that might contain the actual SSID belonging to the password.
The assumption that the sniffed passwords correspond to SSIDs that were also transmitted could additionally be verified by setting up fake access points on the fly using the potential credentials we observed.
As that would constitute an active attack on the devices and since we are determined to improve user security, not undermine it, we decided against employing fake-AP attacks.

Moreover, Wi-Fi locations can often be gathered using Wi-Fi mapping services like WiGLE, as we demonstrate in \cref{sec:data_analysis_wigle}.
Additionally and with enough criminal energy, an attacker could follow the owner of a talkative device to their home and try out the password in their home network.

\subsubsection{Broadcasts of SSID Mistypes.}\label{sec:typos}

In a manual analysis of the data set, we found various devices that broadcast multiple different spellings of presumably the same SSID.
We assume that users manually entered them into their devices while trying out for different spelling and capitalisation variations, e.\,g., \texttt{my~network}, \texttt{MY\_NETWORK}, \texttt{MyNetwork}.
We quantify the amount of mistyped SSIDs by calculating the normalised edit distance between all SSIDs in a burst stemming from a single device. The edit distance defines the minimum amount of operations (insertions, deletions or substitutions of characters) needed to transform one string into another. 
Since the edit distance can also be calculated over strings of different length, we normalise the result with respect to the longer string length, i.\,e., we divide the edit distance by the maximum length of the two SSIDs.
Similar strings have an edit distance close to 0, while the edit distance is closer to 1 the more strings differ. Before calculating the edit distance, all input strings are transformed to lowercase, as otherwise, the normalised edit distance between SSIDs like \texttt{NETWORK} and \texttt{network} would evaluate to 1.
We set the threshold at which the strings are considered similar and thus treated as mistyped to 0.3.
This way, strings that differ in less than 30\,\% of their letters are considered similar. We decided on such a high threshold to accommodate short SSIDs as well as long ones.
A manual inspection verified that the results fulfill the typo criteria and nevertheless do not contain distinct network names like "Fritz!Box 7490" and "Fritz!Box 7590".   

%
We found that 19.9\,\% of the transmitted SSIDs, stemming from 138~distinct bursts, are similar enough to another SSID in the same burst to be considered a typo.
Such a set of constantly transmitted misspelled SSIDs increases the fingerprint of a device drastically and makes tracking it easy.

\subsubsection{Additional findings.}\label{sec:additional findins}

We found at least one string that corresponds to a store and the Wi-Fi password of the store's internal Wi-Fi. We deem this highly likely as it began with the letters "PW:" and contained the name of the store in the password.
We identified 106 distinct first and/or last names, which were propagated 3339 times over the course of the experiment.  
We found three e-mail addresses that were propagated 36 times. 
We identified 92 distinct holiday homes or accommodations whose SSIDs users had added to their list of known networks, which were propagated 1257 times. In addition, we found the name of a local hospital broadcast in two different spelling variations 15 times. 
It is particularly shocking to see such sensitive information like an e-mail address being transmitted openly, let alone the hospital name, from which a potential stay at the hospital can be inferred. At the same time, the name of a person or the hotels and holiday homes in which they have stayed can also be used to draw conclusions about the person. 

\subsection{Geolocation Discoverability and Uniqueness}
\label{sec:data_analysis_wigle}

To provide a better estimate of whether an SSID exists or not, we ran all observed SSIDs trough the geolocation lookup API of WiGLE.
This way, we were able to find out whether the captured SSIDs correspond to actual APs catalogued by WiGLE.
Of course, this approach has one limitation: Mobile devices should, in a perfect scenario, only transmit SSIDs of hidden networks. Those should not be included in the WiGLE map at all, as the service only maps the transmitted SSIDs.

To evaluate the uniqueness of location of the SSIDs we found, we performed an analysis on the coordinates that WiGLE returned.
To reduce the accuracy of the location estimation, we limited the amount of decimal places of the coordinates to 2, thereby providing an approximate 1-kilometre radius in which the actual network can be found.
This also removed artifacts like multiple networks with the same SSID found within a radius of a few metres, that most likely belonged to the same network. Our input consisted of 1478~unique SSIDs. We had to limit our evaluation to a subset of 1440 SSIDs, as the remaining 38 contained special characters, which WiGLE can't resolve. We were able to pinpoint 334 SSIDs to one unique location and 377 SSIDs returned multiple locations. 729~(50.6\,\%) of the SSIDs could not be localised anywhere in the world. The latter are either hidden networks that weren't mapped by WiGLE due to being hidden or mistyped SSIDs. 

\subsubsection{Password Evaluation.}
To provide an estimate of how many SSIDs contained passwords, we filtered the list for 
\begin{itemize}
    \item strings that contain 16 or more numeric digits and
    \item strings that contain "pass", "pw", "kennwort" (the German word for password) or "wpa".
\end{itemize} 
Our input consisted of 77 unique strings classified as passwords. The WiGLE evaluation resolved only a single one of the strings to a unique location. We infer that this is strong evidence that the identified strings are in fact actual passwords.

\subsubsection{Typo Evaluation.}

We performed the same evaluation on the potential typos we identified. In this analysis, we inserted all spelling variations we found of an SSID into our evaluation, which amounted to 296 unique strings. The hypothesis in this case was that at least half of the SSIDs we identified were in fact typos and would not resolve to an existing SSID. We assumed that it would be more than 50\,\% since quite a few of the potential typos had more than one spelling variation. We assumed that the remaining SSIDs would contain the correct and actual spelling that could resolve to an access point. 

Our analysis showed that of the 296 strings classified as potential typos, we were able to resolve approximately 41.9\,\% of the SSIDs and identified 47 unique locations and 66 cases of multiple locations. These results support our hypothesis.

\subsubsection{Limitations.} 
The evaluation of networks contained in WiGLE is severely limited: Only networks that are \textit{not} hidden appear in it, while probe requests of current devices target \textit{only} hidden networks. Nevertheless, we found a large percentage of the networks do in fact exist. 
This can have several explanations: (1) multiple networks with the same name exist, and the one in question is in fact a hidden one, (2) the network has been set to hidden recently, and the map still contains the result of a scan from before, (3) the network was manually added in WiGLE.

\section{Mitigations for Increased Privacy}\label{sec:Privacy Enhancement for the use of Hidden Networks}

Our experiments detailed in the previous section suggest that users, presumably mostly by accident or unwillingly, add items to their list of preferred networks, including credentials and sensitive information.
Together with legitimately added hidden networks, those threaten user privacy by being sent in plain, making this information observable and the users traceable.
To mitigate the issue, we first present a proposal to avoid plain text transmission and then show approaches to limit and control traceable SSIDs through the user interface.

\subsection{Hashing SSIDs in Probe Requests}

Recall that the introduction of the wildcard SSID in probe requests
makes active scanning with specific SSIDs only necessary for hidden networks.
While some publications consider hidden networks obsolete and no longer recommended~\cite{GoovaertsPassiveNotActive}, a recent study~\cite{letNumersTellTheTale} revealed that in some areas, up to 44\,\% of the detected networks were hidden.
While the WPA3 standard contains improvements to the confidentiality management frames~(802.11w)~\cite{PMF-WiFi}, this standard only applies to frames transmitted after a 4-way handshake, and not to frames transmitted without handshakes.
Consequently, probe requests are not protected.
To rectify this, we propose the following mitigation.

\subsubsection{SSID Hashing.}
To circumvent the need to send cleartext SSIDs, we propose to send them in a hashed and salted manner instead. The device emitting the probe request would first salt the SSID using its randomised sender MAC address and the sequence number of the packet and then hash it. It would then send the hash, but omit both salt components as they are included in the frame anyway like so:
\begin{displaymath}
  send(hash(MAC || SN || SSID))
\end{displaymath}

Access points of hidden networks would then, upon receiving the probe request, prepend the MAC address and sequence number as salt to their own SSID, hash it, and compare the result with the received hash. If they match, the client was probing for their hidden network.
As the MAC address should be chosen randomly and the sequence number changes with every packet, they introduce sufficient entropy and variability in combination to be suitable as salt to make sure the sent information can not be used to track and identify devices through a constant hash value.
This mitigation could be employed regardless of whether or not a connection to the network has ever been established before, as it does not require the previous exchange of secrets.
Another advantage of this mitigation is that potentially sensitive SSIDs (e.\,g., containing names or passwords) can only be distinguished from other SSIDs (e.\,g., generic home router names) with significant effort of brute-forcing the cleartext, thereby improving the privacy of clients and AP operators.

\subsubsection{Attack Model.}
We consider an attacker that can monitor all probe requests and has bounded computational power, which makes it impractical for her to find preimages~(SSIDs) of the hashes she observes.
Introducing the combined MAC address and sequence number salt makes pre-computing hashes impractical,
provided that MAC addresses are randomised and sequence numbers are also randomly chosen within their full 12~bit value range.
If the randomised MAC address contributes 24~bit of entropy, leaving out the EUI/CID parts as a lower estimate,
this totals to 36~bit of salt entropy.
Attackers with a priori knowledge of used SSIDs can however, reduce costs significantly compared to true brute-forcing, depending on the number of likely used SSIDs.
Consequently, tracking a device whose SSID set is known is therefore feasible.

To determine the practical feasibility of a hash-based mitigation, we first evaluate the computational cost and then calculate the additional bandwidth requirement. 

\subsubsection{Computational Overhead.}
Using the SHA-256 hash from Python's \texttt{hashlib} library, we prototyped the hashing and comparison that an AP would have to perform: first hashing a received SSID and then comparing it to a known hash (the AP's SSID).
For baseline reference, we implemented the current string comparison of SSIDs.
We performed one million computations in three runs on a Raspberry Pi Model~3, which is likely a good lower estimate of computational power of most routers.
A single computation with hashing required on average 11.7~microseconds pure CPU time, while the baseline required an average of 4.6~microseconds.
Using hashing therefore increases the time by 153.7\,\%.
Considering that the Raspberry Pi does not have cryptographic hardware acceleration, which professional Wi-Fi routers might have, we additionally ran the experiment on a laptop with an Intel i5, where hashing only added an average overhead of 53\,\%.

In our experiment in a busy pedestrian zone, we captured around 23~probe requests per second, of which only 23.2\,\% contained an SSID\@.
Therefore, following these figures, an AP would have to hash approximately 5.3~probe requests per second.
As a Raspberry Pi can perform around \num{85200}~hashing operations per second,
we deduce that hashing and comparing should be well within the available resources for similarly equipped APs even in much more frequented deployment locations.

\subsubsection{Bandwidth Overhead.}\label{Bandwidth Overhead}

Our proposal may also introduce a bandwidth overhead by always occupying
the full \SI{32}{bytes} available for SSIDs in a probe request frame~\cite[Sec. 9.4.2.2]{WLAN-spec},
whose length would otherwise vary with the actual SSID length.
The average length of all packets in our city centre capture is \SI{133.3}{bytes}, while the average length of packets containing SSIDs is \SI{147.0}{bytes}.
In our capture, the average length of SSIDs was \SI{11.4}{bytes}.
If all SSIDs were transmitted as hashes, it would increase the size by \SI{20.6}{bytes}, leading to an average size of packets with SSID of \SI{167.6}{bytes}, which is an increase of 14.01\,\%. 
%
Considering that probe requests make up a tiny fraction of the actual transmitted traffic, we consider this an acceptable trade-off for more privacy and less fingerprintability.


\subsection{Mitigations through User Interface Design}\label{sec:evaluation}

\begin{figure}[t]
\begin{subfigure}[t]{.66\textwidth}
    \centering
    \includegraphics[scale=0.149]{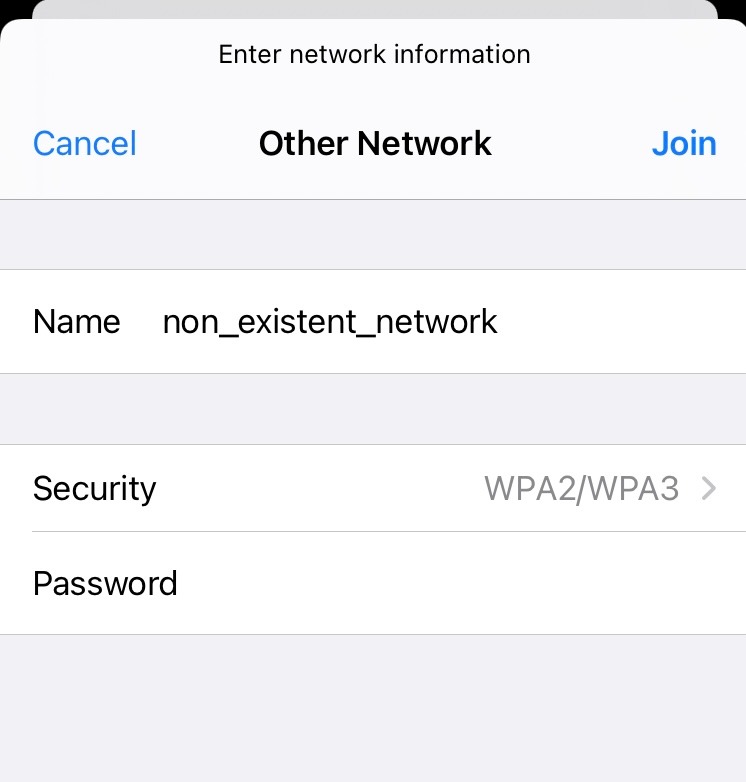}
    \hfill
    \includegraphics[scale=0.15]{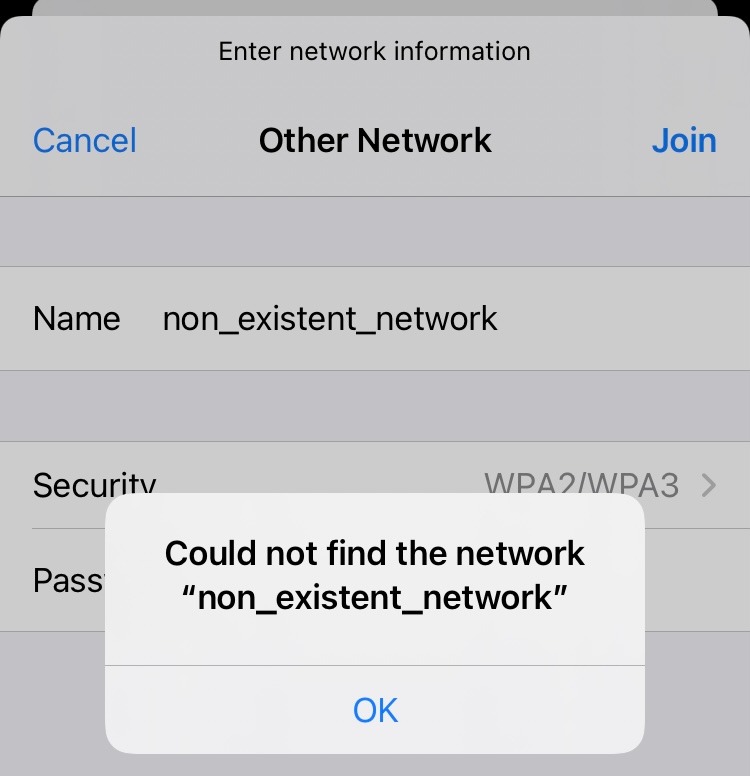}
    \caption{Attempting a connection to a hidden network on iOS 15. Contrary to Android devices, iPhones only allow to manually add networks to which a connection can be established.}
    \label{fig:failed_network_conn}
\end{subfigure}
\hfill
\begin{subfigure}[t]{.32\textwidth}
    \centering
    \includegraphics[scale=0.13,trim=0 1.3cm 0 6.5cm, clip]{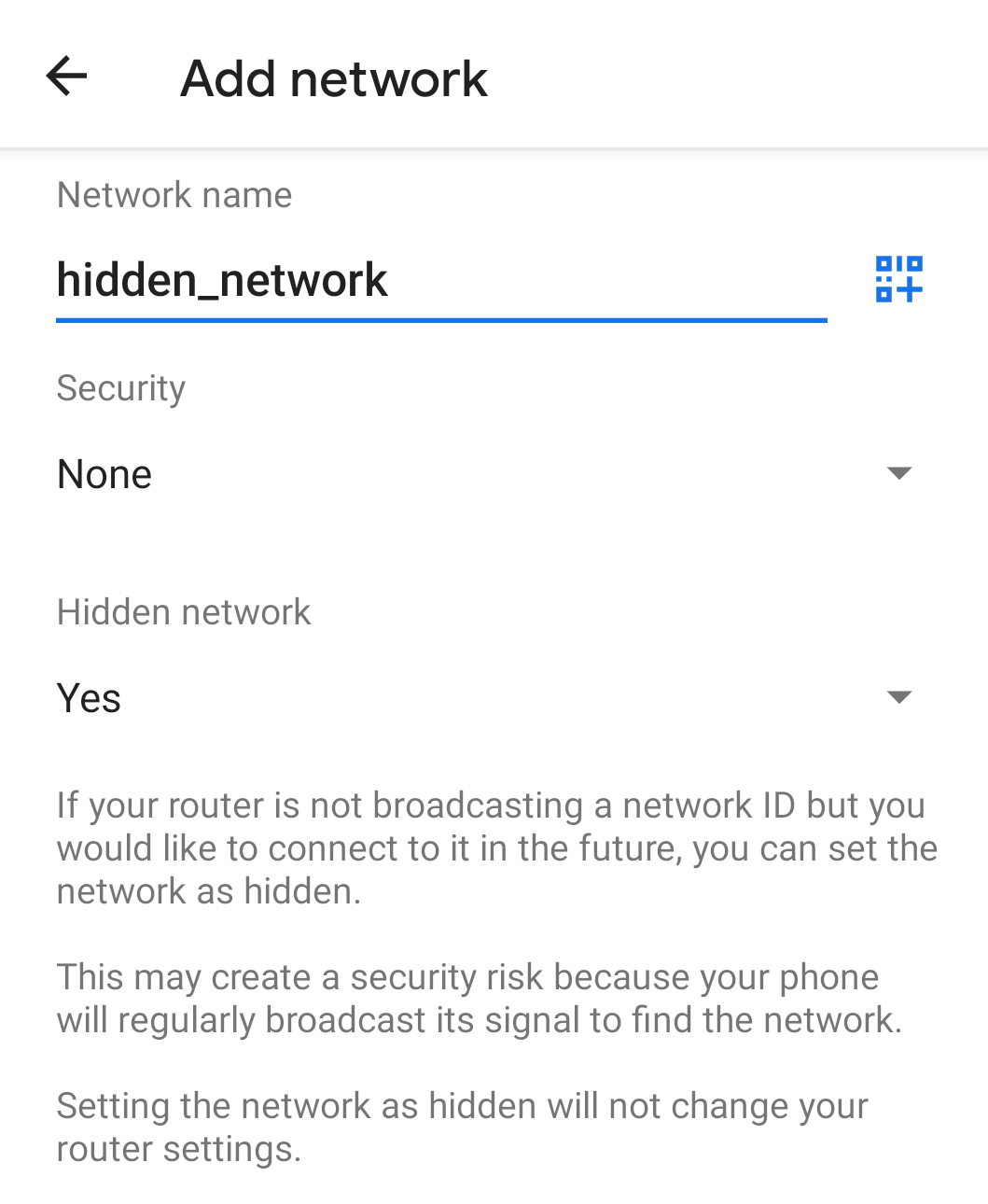}
    \caption{Warning message when adding a hidden network on Android 9 or newer.}
    \label{fig:hidden_message}
\end{subfigure}
\caption{User dialogues mitigating unwanted SSID entry.}
    \label{fig:screenshots}
\end{figure}

Current iOS and Android version already employ mechanisms to prevent users from accidentally adding items to their PNL, to maintain that list and to change the connection behaviour for individual networks on this list.
In the following, we briefly summarise the status quo and suggest further improvements for more accessible and effective controls.

\subsubsection{SSID Entry Safeguards.}
Both iOS and Android have safeguards against accidentally adding hidden networks: iOS will only add networks to which a connection can be established at the time of entry (see \cref{fig:failed_network_conn}).
In Android, a manually added network is no longer automatically considered a hidden network. Instead, to enter a hidden network, users have to explicitly select it and then receive a warning about the privacy risks (see \cref{fig:hidden_message}).
We suggest to combine both measures for manually adding SSIDs.

\subsubsection{Known SSID Removal.}\label{sec:known_ssid_removal}
Being able to remove entries from the list of preferred networks should be possible to reduce traceability and susceptibility for fake AP attacks~\cite{Vanhoef-RandomisationNotEnough}. 
However, as mentioned in \cref{sec:Differences between Android and iOS versions}, removing a known network from proximity is not straightforward in iOS. 
On Android, in contrast, the list of known networks can be modified directly and at all times.

\subsubsection{PNL Entry Expiry.}
In order to maintain its usability,
user-facing lists of preferred networks should implement measures against cluttering, which result from steadily adding new SSIDs over time.
To avoid cluttering with no longer needed SSIDs, e.\,g., added during temporary stays, we propose an expiration date for SSID entries: Upon adding a new SSID, the user is prompted to choose when the SSID should be forgotten again.
The default expiry date would be \textit{never}, but for SSIDs that are knowingly only in use for a limited period of a few days or weeks, users can choose accordingly.
By limiting the life span of a network entry, its negative effects on traceability and the chance to exploit them using fake AP attacks are at least temporally limited.
To the best of our knowledge, there is no such mechanism in any OS, but to implement it could significantly reduce the amount of SSIDs accumulated over time.

\subsubsection{Adjustable Auto-Joining.}
In addition to networks of limited temporal relevance that benefit from expiry,
there are also networks users use only occasionally but regularly, e.\,g., once every few months or years.
While it is convenient to keep them in the PNL, this again increases the risk of fake AP attacks.
Instead of removing such networks, preventing the automatic connection to them already effectively reduces the fake AP attack risk.
Both iOS and Android offer ways to disable auto-joining on a per network basis.
To make this option more visible and broadly realise privacy gains, we suggest to again prompt users when initially joining a network, whether they want to automatically or manually connect in the future.

\subsubsection{Silencing Probe Requests.}
For particularly high privacy demands, disabling probe requests altogether might be an acceptable trade-off.
We therefore suggest an advanced network setting, where users are able to choose that their devices do not send active probe requests \emph{at all}, knowing that (a) connection establishment relies only on passive AP announcements and might be slower, (b) the battery usage might be higher, and (c) the connection to hidden networks would be impossible.
Such behaviour could also be part of a \emph{reduced visibility} mode that user, e.\,g., activate in the control center, when they pass through an untrusted area like a shopping centre known for extensive visitor analytics, similar to a do-not-disturb switch.

\section{Discussion}
\label{discussion and limitations}

\subsubsection{Legal Consideration of SSIDs.}
As mentioned in the introduction, Wi-Fi tracking is also used to measure pedestrian flow and count in cities.
In two German cities, such measurements have been conducted until the responsible authority started investigations~\cite{datenschutzzentrum}.
The authority's reasoning was that MAC addresses are personal data and it is therefore not legal to record them without legal basis.
In both cases, the measurements were ceased.
Especially as the continued roll-out of MAC address randomisation might give renewed support to the legal positions that (randomised) MAC addresses should no longer be considered identifying information,
we argue that this assessment can and should not be limited to the sender address of probe requests.
Considering the wealth of personal and sensitive information we observed in SSID fields,
they can constitute identifying information as well and thus require due consideration.
For instance, for 334 of the SSIDs we measured, we were able to identify a unique location the SSID originated from.
We therefore argue that at least for as long as there are still devices broadcasting SSIDs, probe requests should be considered personal data and not be used for monitoring without legal basis. 

\subsubsection{Intentional Password Broadcast.}
Some APs might intentionally broadcast their password as SSID to allow visitors an unrestricted yet encrypted Wi-Fi connection,
as an alternative to unprotected Wi-Fi. It is a limitation of our password leakage evaluation that we cannot distinguish those intentional from unintentional cases.
However, the fact that we could only resolve one of these SSIDs in our geo-lookup suggest that the ratio of SSIDs to actual APs is rather low. Additionally, the notion of intentionally storing passwords in the SSID field should have been superseded by the introduction of OWE (Opportunistic Wireless Encryption): OWE describes the unauthenticated but encrypted connection between two devices, and is contained in the Wi-Fi specification under the name \textit{Wi-Fi CERTIFIED Enhanced Open™}\cite{EnhancedOpen}.

\subsubsection{Deployment of Hashed SSID Scheme.}

Our proposed hash-based scheme requires modifications to the Wi-Fi implementations: The mobile device has to apply the hashing algorithm to its temporary MAC address, sequence number, and the SSID of the sought-after network before transmission.
The AP, upon receiving a probe request, has to apply hashing to the MAC address and sequence number as salt to its own SSID. On one hand, these changes are easy to implement, on the other hand they require a widespread deployment to be effective. While the privacy gain would be worth the deployment effort, it is unfortunately likely that only newer devices would profit from the scheme, while older devices would remain unpatched.  

\subsubsection{Limiting Bandwidth Overhead.}

Our proposed method of salting and hashing the SSIDs of hidden networks to improve user privacy introduces a bandwidth overhead, (cf.\ \cref{Bandwidth Overhead}):  The average length of a packet containing an SSID would be increased by 14\,\%. This could be addressed by truncating the hash to, e.\,g., \SI{16}{bytes} before inserting it into the SSID field.
That way, the average packet length would be reduced to \SI{151.6}{bytes}, which results in an overhead of only 3.2\,\%. At the same time, this reduces the security of the system, as hash collisions become more likely.

\subsubsection{Impact of OS Support Lifespans.}
A contributor of non-wildcard probe requests in the wild, besides hidden networks,
are legacy devices.
While devices running Android~10 or newer use MAC address randomisation and omit SSIDs, older devices do not receive updates long enough to benefit from this improvement.
This especially disadvantages people with budgetary constraints or sustainability in mind, who keep their devices longer.
This can only be rectified by longer support lifespans that should be legally mandated if manufacturers do not move voluntarily.
Such changes with significance to user privacy should be considered like critical security patches and be back-ported to older versions.

\section{Conclusion}
\label{conclusion}
Probe requests are plainly observable to everyone around a sending device. Since they can contain sensitive data, they should be sent more carefully and with privacy in mind. 
We have collected and analyzed data in a pedestrian zone to gather insight into the status quo of probe requests in the wild. We identified a wealth of personal information in transmitted SSIDs such as potential passwords, holiday homes and e-mail addresses, which should be considered personal and sensitive data. For 334~SSIDs, we were even able to resolve their unique geographic location. 

In the area of anonymisation of probe requests, some progress has been made in the last few years, with the latest mobile operating system updates. Nevertheless, we are still facing problems in terms of user privacy. 
To minimise the amount of personal data that can be sent accidentally, we propose mitigations for both the network layer and the user interface. The first consists of hash-based concealment of SSIDs using a salt constructed from the MAC address and the sequence number of the request. To demonstrate its feasibility we provide estimates of the computational overhead and additional bandwidth requirements. For the latter, we propose to remodel the Wi-Fi handling of mobile devices to add only existing networks in range and to empower their users to take more control over their PNL and minimise the amount of potentially exploitable data. 

\section*{Acknowledgements}

We would like to thank our reviewers for their valuable and constructive feedback.

\appendix
\section{Appendix}
\label{sec:appendix}

\subsection{Ethical collection of probe requests}
\label{sec:Ethical collection of probe requests}

Following approval and in coordination with the ethics committee of the informatics faculty of the University of Hamburg, we conformed to the following measures to observe and protect users' privacy rights: 
\begin{itemize}
    \item During the time of the experiment, we set up a well visible sign declaring the undergoing probe request monitoring, including information on how to contact the person in charge.
    \item We informed and obtained consent from building management to conduct the experiment.
    \item We provided an option to remove recorded probe requests should participants state their non-consent.
    \item We used off-the-shelf wireless USB antennae with a limited range to narrow the radius of our measurement 
    \item In case the data set contains personal information, we either anonymise it before storing, or delete it directly after analysing it.
    \item Any personal data is stored securely, both technically as well as organisationally, to prevent misuse.
    \item To preserve location privacy, we limit the amount of decimal places of the coordinates returned by WiGLE to 2, thereby providing an approximate 1-kilometre radius in which the actual network can be found.
\end{itemize}

\bibliographystyle{splncs04}
\bibliography{sources}

\end{document}